# LIFETIME OF HIGHLY EFFICIENT H- ION SOURCES*

V. Dudnikov#, Muons, Batavia, IL 60510, USA
D. Bollinger, FNAL, Batavia, IL 60510, USA, D. Faircloth, RAL, Oxfordshire, OX11 0QX, UK,
S. Lawrie, JAI AS, Oxford, UK and ISIS, RAL, Oxfordshire, OX11 0QX, UK

*Abstract*

Factors limiting the operating lifetime of Compact Surface Plasma Sources (CSPS) are analyzed and possible treatments for lifetime enhancement are considered. Noiseless discharges with lower gas and cesium densities are produced in experiments with modified discharge cells. With these discharge cells it is possible to increase the emission aperture and extract the same beam with a lower discharge current and with correspondingly increased source lifetime. A design of an advanced CSPS is presented.

## INTRODUCTION

Compact Surface Plasma Sources (CSPS) [1-3] such as magnetron, semiplanotron, Penning Discharge (PD) SPS can have high plasma density (up to $10^{14}$ cm$^{-3}$), high emission current density of negative ions (up to 8 A/cm$^2$), have small (1–5 mm) gaps between cathode emitter and a small extraction aperture in the anode. They are very simple, have high energy efficiency up to 100 mA/kW of discharge (~100 times higher than a modern large Volume RF SPS [1-3]) and have a high gas efficiency (up to 30%) using pulsed valves. CSPSs are very good for pulsed operation but electrode power density is often too high for dc operation. However, CSPS were successfully adopted for DC operation with emission current density j~300 mA/cm$^2$ in hollow cathode Penning discharge SPS and in spherical focusing semiplanotron SPS. Flakes from electrode sputtering and blistering induced by the discharge and by back accelerated positive ions are the main reasons for ion source failure. Suppression of back accelerated positive ions and flake evaporation by pulsed discharge can be used to significantly increase the operating lifetime of CSPS. Noiseless discharges with lower gas and cesium densities are produced in modified discharge cells. With these discharge cells it is possible to increase the size of the emission aperture and extract the same beam from a lower current discharge with a corresponding increase source lifertime. Design of advanced CSPS is presented. Extrapolated H- beam parameters are up to 100 mA pulsed current and up to 15 mA average current with lifetime up to $10^3$ hours (10 A hours).

The operational ISIS PD SPS [2] has very small discharge cell (5x3x11 mm$^3$) and for noiseless discharge production it is necessary to use a high gas and cesium density. A narrow 0.6 mm emission slit is needed to prevent extraction voltage arcing. Very high emission current densities of 1.5 A/cm$^2$ and high discharge current densities of J~150 A/cm$^2$ are destructive for discharge cell electrodes. However, it can operate for up to 48 days (1200 hours) with pulsed beam current ~50 mA after bending magnet with discharge duty factor 2.5%.

To use a more relaxed discharge parameter and corresponding increased lifetime it is necessary to produce the noiseless discharges with low gas and cesium density (like in large volume SPS [1]). Experiments with modified discharge cells for noiseless discharge production are described below.

## NOISELESS DISCHARGE WITH MODIFIED DISCHARGE CELLS

In ISIS PD SPS the cathodes is deposited by Molybdenum, sputtered from the anode insert and replaced by new one after every source cleaning. A big stock of used cathodes is accumulated. To explore the possibility to reuse the old cathodes, a small processing modification was proposed to the ISIS PD SPS. Namely, the damaged parts of cathode were removed and the plasma plate thickness increased for better cooling of the H- emitting area around the emission slit.

Several ISIS PD SPS cathodes were sent from RAL to Fermilab and processed as shown in Fig. 1.

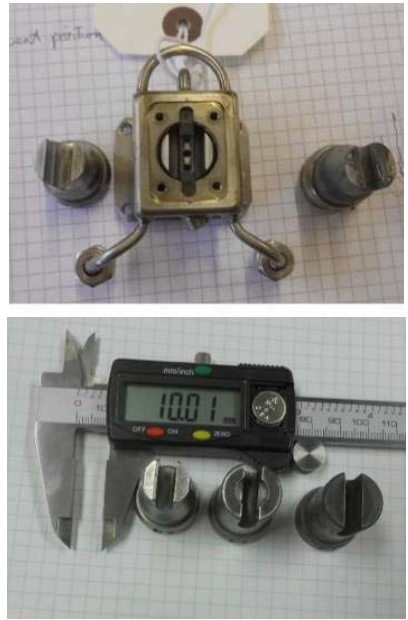

Figure 1: Modification of ISIS PD SPS: (Top) right- standard cathode with a discharge gap d=5 mm; left- modified cathode with gap d=7 mm; central-PD SPS assembling with modified cathode gap d=9 mm and the

*Work supported by grant DE-SC0006267, and STFC JAI grant ST/G008531
#Vadim@muonsinc.com
Operated by Fermi Research Alliance, LLC under Contract No. De-AC02-07CH11359 with the United States Department of Energy.

cathode modification with a drilling a cylindrical extension of the working cathode gap (bottom).

The cathode was assembled with the PD SPS body as shown in Fig. 1 (top) and the discharge and beam extraction were tested in the test stand. As a first step, a discharge supported by a DC power supply with hydrogen injection by piezoelectric valve with frequency of 50 Hz was tested and used for PD SPS conditioning and cesiation.

Pulses of discharge current starting at voltage Ud=600 V are shown in Figs. 2-4. The discharge current is high at high gas density and decreases with decreasing gas density until the discharge stops.

Fig. 2 shows the discharge current signal in the PD with a standard gap of d=5 mm, with DC voltage Ud=600 V and pulsed gas 50 Hz. The discharge is noisy during the entire pulse with the standard conditions: vacuum p=7.3 $10^{-5}$ Torr; magnetic coil current Im=13 A.

This experiment shows that in a standard cell with a 5 mm cathode gap, the discharge is noisy and a special attempt for noiseless discharge production is needed by increasing gas flow and volume cesium density.

Fig. 3 shows the discharge current in the PD with gap d=7 mm, with DC voltage Ud=600 V, pulsed gas 50 Hz. The discharge is noiseless until the gas density decreases near the end of the discharge. (same standard conditions: vacuum p=7.3 $10^{-5}$ Torr; magnetic coil current Im=13).

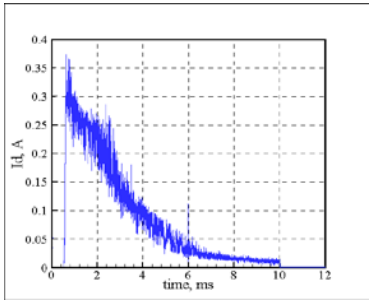

Figure 2: Discharge current in PD with gap d=5 mm, with DC voltage Ud=600 V, pulsed gas 50 Hz.

The discharge is noiseless for ~2 ms, which is very favourable for high brightness H- beam production at low cesium density, which is favourable for stable extended operation.

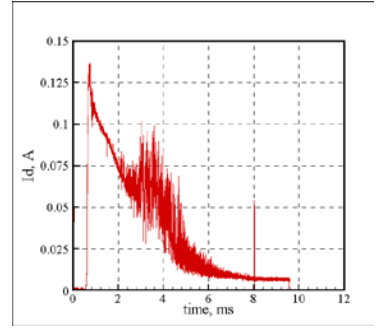

Figure 3: Discharge current in PD with d=7 mm cathode gap . Noiseless discharge achieved for ~2 ms.

By the increasing the cathode gap to 9 mm, the discharge stability was further improved as shown in Fig. 4. With the same operation conditions the discharge current is noiseless for an even longer duration until the gas density decreases near the end of discharge.

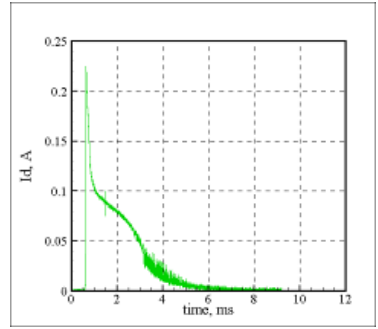

Figure 4: Discharge current in PD with d=9 mm cathode gap. Noiseless discharge achieved for ~3 ms.

The discharge stability in the cell with d=7 mm cathode gap was improved by increasing the gas density. The noise generation increases the discharge current as well the increase the gas density because transverse electron mobility in a magnetic field increases with increased scattering frequency.

The dependence of the effective transverse electron mobility μ on the scattering frequency ν and the electron cyclotron-frequency ω is expressed by:

$$\mu = e\nu/m\,(\nu^2 + \omega^2) \qquad (1)$$

and is shown in Fig. 5 (bottom). The effective transverse electron mobility μ increases at low scattering frequency ν below the cyclotron frequency ω and decreases at higher ν. Transverse mobility can be increased through electron scattering by plasma fluctuations connected with plasma turbulence. For this reason the plasma instability development is thermodynamically "profitable at low gas density and strong magnetic field and "non-profitable" at high gas density n and low magnetic field B.

A diagram of the magnetron discharge stability as a function of magnetic field B and gas density n is shown in Fig. 5 (top). The diagram for Penning discharges has a similar shape with different parameters. For discharge triggering at low gas density, the magnetic field B is used to prevent direct collection by the anode of electrons emitted by the cathode. Higher magnetic field is necessary for lower gas density. The boundary of gas discharge triggering in the top diagram B, n of Fig. 5 can be presented by expression:

$$(n-n_{min}) \times (B-B_{min}) = C, \qquad (2)$$

where C depends on the discharge cell configuration. In this diagram, $B_{min}$ is the smallest magnetic field necessary for discharge triggering. The gas density n also should be higher than $n_n$. Gas density for different discharge conditions can be calibrated using a measured gas density supported in the vacuum chambers with ion source.

At gas density below a critical density n* it is possible to have only a noisy discharge. At gas densities above n* the discharge becomes noisy at high magnetic field and noiseless at lower B. For reliable ion beam extraction, the gas density in the discharge cell should be lower than the level $n_m$ at which the probability of extraction voltage breakdown becomes high or H- stripping become significant. With higher $n_m$ it is need to use a small emission aperture. With slit emission aperture it is possible to have a larger total emission area because the gas target thickness after the emission aperture is determined by the small dimension of the slit and higher perveance for beam extraction.

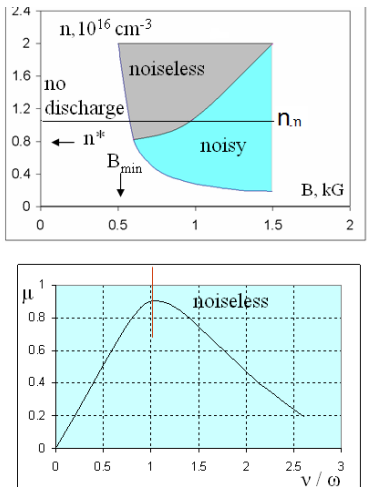

Figure 5: Diagram of the magnetron and Penning discharge stability as a function of magnetic field B and gas density n (top) and the dependence of the effective transverse electron mobility μ on the ratio of scattering frequency ν to cyclotron frequency ω (bottom).

For the standard ISIS PD SPS discharge cell with d=5 mm cathode gap and anode window hxl=2x11 mm$^2$ shown in Fig. 1 (right), the condition for normal beam extraction corresponds to a vacuum gauge reading p=7.3 10$^{-5}$ Torr and a magnetic coil current of Im=13 A.

With these parameters, the discharge without cesium is noisy during the entire gas pulse as shown in Fig. 2. In this discharge cell the noiseless discharge can be produced only at the end of the high current discharge pulse with a high source body temperature of Tb~460 C. For this cell, $n_m$ <n* for hydrogen and only adding a large density of cesium into the volume shifts n* below $n_m$ and permits noiseless discharge production.

This circumstance significantly increases cesium consumption, increases electrodes sputtering by back accelerated cesium ions, decreases the efficiency of negative ion production, and decreases ion source lifetime several times.

Discharge properties changed dramatically by increasing the cathode gap to d=7 mm. With the same standard discharge conditions, p=7.3 10$^{-5}$ Torr, Im =13 A, the modified discharge cell with cathode shown in Fig. 1 (left), the discharge is noiseless without cesium during the first 2 ms when the gas density is high enough. In this situation there is no need to have cesium in the cell volume and one only needs to have half of a monolayer of cesium film on the plasma electrode plate around the emission aperture for enhanced H- production A noiseless beam with good emittance and high brightness can then be extracted from the entire discharge pulse. This decreases cesium consumption, decreases electrodes sputtering, increases efficiency of H- generation and extraction, and should increase SPS lifetime up to several times. Increasing the cathode gap to d=9 mm as shown in Fig. 1 (middle), the discharge stability was further improved as shown in Fig. 4. The discharge without cesium is noiseless for 53 ms and the gas density can be decreased without loss of the noiseless part of the discharge. In terms of expression (2), the increase of the cathode gap decreases the minimal gas density $n_{min}$ and minimal magnetic field $B_{min}$, necessary for triggering the Penning discharge. At this the gas density, n* is shifted below $n_m$ and the large noiseless area for beam production becomes available.

With lowered gas and cesium density in the discharge cell it is possible to increase the emission slit width and produce the required beam current at a lower discharge current and increase the SPS lifetime significantly.

The possibility to increase the emission slit was tested using scaled versions of a PD SPS at LANL [2]. The transverse normalized emittance was below 0.2 $\pi$ mm-mr with the emission slit width increased up to 2.8 mm. With such increase of the slit width it is possible to produce the necessary beam current with a discharge current up to 5 time less than with the existing 0.6x10 mm$^2$ slit. In the scaled versions of a PD SPS at LANL larger anode windows were needed, meaning they had to increase the discharge current, which decreased the efficiency of negative ion generation. A small increase of anode window size up to 3x15 mm$^2$ would have been useful for easy discharge triggering and reducing the necessary gas density, yet without the decrease of the H$^-$ generation efficiency.

## SUMMARY

Optimization of the discharge cells in a Penning H$^-$ ion source is a viable method for increasing the phase space of the stable region for noiseless discharge production. With this method, cesium usage would be decreased, potentially resulting in longer source lifetimes.